# DIPOLE-QUADRUPOLE THEORY OF SURFACE ENHANCED INFRARED ABSORPTION AND APPEARANCE OF FORBIDDEN LINES IN THE SEIRA SPECTRA OF SYMMETRICAL MOLECULES

A.M. Polubotko

A.F. Ioffe Physico-Technical Institute Russian Academy of Sciences, Politechnicheskaya 26

194021, Saint Petersburg Russia, Tel: (812) 292-71-73, Fax: (812) 292-10-17,

E-mail: alex.marina@mail.ioffe.ru

## Abstract

The paper presents main aspects of the Dipole-Quadrupole theory of Surface Enhanced Infrared Absorption (SEIRA). It is pointed out the possibility of appearance of the lines, caused by totally symmetric vibrations transforming after the unit irreducible representation, which are forbidden in usual infrared absorption spectra in molecules with sufficiently high symmetry. Observation of such lines in the SEIRA spectra of diprotonated $BiPyH_2^{2+}$ and ethylene, adsorbed on $Cu$ and on mordenites is pointed out. The results well agree with our ideas about surface enhanced optical processes, based on the conception of a strong quadrupole light-molecule interaction, which allows us to develop the SERS and SEHRS theories.

## Introduction

The theory of enhancement of optical processes by molecules, adsorbed on rough metal surfaces is of great interest. For a long time and now many investigators assign these effects to the enhancement of the electric field due to excitation of plasmons on rough metal surfaces. In addition, one of the common conceptions used by researchers is the one of chemical enhancement, which is associated with charge transfer from the surface to the molecule. Another explanation of these processes, which is sufficiently widespread in literature is the one, based on the rod effect, or on the increase of the electric field near such a roughness as a cone or a spike. This explanation was used in our works. However additionally we pointed out the strong increase of the quadrupole interaction in such system. This theory is named as the Dipole-Quadrupole theory. At present it allows us to explain completely such phenomena as SERS [1] and SEHRS [2-5].

It was demonstrated in [6] that SERS and SEIRA have similar regularities. In [7] we demonstrated that the Dipole-Quadrupole theory is able to explain the magnitude of the enhancement in SEIRA. In addition it was demonstrated that the enhancement due to the quadrupole interaction can be stronger than the one, caused by the dipole interaction. One of the most important regularities, which arise in the surface enhanced optical spectra, is appearance of forbidden lines. In [8-10, 2-5] it was demonstrated that the quadrupole interaction allows to explain successfully their appearance in SERS, SEHRS and SEIRA. In [8] we predicted appearance of forbidden lines in SEIRA and obtained selection rules for the contributions in the scattering. However the forbidden lines were not investigated experimentally in these processes for a long time. The main reason was that investigators did not measure the SEIRA spectra on molecules with sufficiently high symmetry, where these lines should be observed. Here we demonstrate that the forbidden lines, predicted in [8] are observed in the SEIRA spectra of ethylene and diprotonated $BiPyH_2^{2+}$, adsorbed on copper and also in the SEIRA spectra of ethylene, adsorbed on mordenites that confirm correctness of our theory.

# Main relations of the Dipole-Quadrupole theory of SEIRA

The enhancement in SEIRA arises due to enhancement of the electric field near rough metal surface, such as in SERS and SEHRS. Good models of roughness, which describe the enhancement of the electric field are a cone or a spike. More detailed description of the theory of electromagnetic field near the rough metal surface one can find in [1]. For the models of the roughness of the cone, or spike forms the radial component of the electric field has the form

$$E_r \sim E_{0,inc} C_0 \left(\frac{l_1}{r}\right)^\beta , \qquad (1)$$

where $E_{0,inc}$ is an amplitude of the incident field, $C_0 \sim 1$ - is a numerical coefficient, $l_1$ - is the characteristic size of the cone or spike, $0 < \beta < 1$, $r$ - is the length of the radius vector with origin on the spike top. The behavior of the field is singular that means its strong enhancement. In addition, the derivatives of the field $\frac{\partial E_\alpha}{\partial x_\alpha}$ or $\frac{\partial E_r}{\partial r}$, in a conical coordinate system, are very large that cause increase of the quadrupole interaction. The expression for the absorption cross-section of symmetrical molecule with the dipole and quadrupole interaction taken into account has the form [7, 8]

$$(\sigma_{abs})_{surf} = \frac{\pi \omega_{inc}}{hc\varepsilon_0} \frac{|\bar{E}_{inc}|^2_{surf}}{|\bar{E}_{inc}|^2_{vol}} \times \sum_p \frac{V_{(s,p)}+1}{2} \times$$

$$\times \left| \sum_i (P^e[d_{e_i}] - \Delta d_{n,(s,p),i}) e_{inc,i} + \sum_{i,k} (P^e[Q_{e,ik}] - \Delta Q_{n,(s,p),i,k}) \times \frac{1}{2|\bar{E}_{inc}|_{surf}} \frac{\partial E_i}{\partial x_k}\bigg|_{surf} \right|^2$$

(2)

Here $\omega_{inc}$ -is the frequency of the incident light, $(\overline{E}_{inc})_{vol}$ - is the incident field in a free space, $(\overline{E}_{inc})_{surf}$ - is the surface field, which affect the molecule, $V_{(s,p)}$ -is the vibrational quantum number of the $(s, p)$ vibration. Here $s$ designates the group of degenerate vibrational states, $p$ numerates the states inside the group.

$$P^e[f] = \sum_{\substack{l \\ l \neq n}} 2\,Re \frac{R_{nl(s,p)}\langle n|f|l\rangle}{(E_n^{(0)} - E_l^{(0)})} \quad . \tag{3}$$

Here indices $n$ and $l$ refer to the electron ground and excited states respectively, $E_n^{(0)}$ and $E_l^{(0)}$ are their energies in the model of motionless nuclei [1, 11]. $f$ designates the components of the dipole $d_{e,i}$ or quadrupole $Q_{e,i,k}$ moments of electrons,

$$R_{n,l,(s,p)} = \sum_{J,i} \sqrt{\frac{\hbar}{\omega_{(s,p)}}} X_{J,(s,p),i} \frac{\partial\langle l|\hat{H}_{e-n}|n\rangle}{\partial X_{J,i}}\bigg|_{\Delta X \equiv 0} \tag{4}$$

is the excitation coefficient of the state $l$ from the ground state $n$ by the $(s, p)$ vibrational mode. Here $X_{J,(s,p),i}$ -is the $i$-th component of the displacement vector of the $J$ nucleus in the $(s, p)$ vibrational mode, $X_{J,i}$ is the $i$ component of the radius vector of the $J$ nucleus. $\hat{H}_{e-n}$ is the Hamiltonian of interaction of nuclei and electrons [1]. The values

$$\Delta d_{n,(s,p),i} = \sum_J eF_J M_J \sqrt{\frac{\hbar}{\omega_{(s,p)}}} X_{J,(s,p),i} \tag{5}$$

$$\Delta Q_{n,(s,p)i,k} = \sum_J eF_J M_J \sqrt{\frac{\hbar}{\omega_{(s,p)}}} (X_{J,i}^0 X_{J,(s,p),k} + X_{J,k}^0 X_{J,(s,p),i}) \tag{6}$$

are deviations of the dipole and quadrupole moments of nuclei from the equilibrium values in the $(s, p)$ vibrational mode.

$$F_J = \frac{Z_J^*}{M_J} \tag{7}$$

where $eZ_J^*$ is the charge, $M_J$ is the mass of the $J$ nucleus. $X_{J,i}^0$ и $X_{J,k}^0$ are the $i$ and $k$ components of the radius vector of the $J$ nucleus in the equilibrium position. One can see from (2) that the SEIRA cross-section is determined by the processes of two types.

1. The processes with the electron shell excitation taken into account, which are described by the terms $P^e[f]$ and are named by the processes of the first type.

2. The processes, caused by the direct interaction of radiation with nuclei, which are described by the terms $\Delta d_{n,(s,p),i}$ and $\Delta Q_{n,(s,p),i,k}$ in [7, 8].

The relative role of these processes is not clear. However it was demonstrated in [7] that the enhancement of the cross-section occurs even in the case, when it is determined by one of these processes only. For the cross-section, caused by the process of the first type [7], the enhancement coefficient is:

$$G_{abs}^e = \frac{(\sigma_{abs})_{surf}}{(\sigma_{abs})_{vol}} \sim C_0^2 \beta^2 \left(\frac{1}{2} B_e\right)^2 \left(\frac{l_1}{r}\right)^{2\beta} \left(\frac{a}{r}\right)^2 , \tag{8}$$

where

$$B_e a \sim \frac{\langle n|Q_{e,i,i}|l\rangle}{\langle n|d_{e,i}|l\rangle} \tag{9}$$

is the ratio of some average values of matrix elements of the quadrupole moments of electrons $Q_{e,i,i}$ with the same indices and of the dipole moments $d_{e,i}$, $a$ - is the molecule size. Here we make averaging on index $i$. Further we consider that

$$\overline{\langle n|Q_{e,i,i}|l\rangle} \sim \overline{\langle n|Q_{e,i,i}|n\rangle} \sim \overline{Q}_{n,i,i} , \tag{10}$$

where $\overline{Q}_{n,i,i}$ is of the order of magnitude of the quadrupole moments of nuclei, while

$$\overline{\langle n|d_{e,i}|l\rangle} \sim \sqrt{e^2\hbar/2m\omega_{n,l}} \times \sqrt{\overline{f}_{n,l}} , \qquad (11)$$

is expressed via some mean value of the oscillator strength $\overline{f}_{n,l}$. From formulae (8-11) one can see that the larger the molecule, the larger the enhancement coefficient. For the molecule $KCN$ we obtained the enhancement coefficient $G^e_{abs} \sim 50$ for reasonable values of the parameters in [8]. However for the same values of the parameters ($C_0 \sim 1, \beta \sim 1, l_1 = 10nm$, $a \sim 0.5nm$, $r \sim 2-2.5nm$) and $B_e \sim 165$ that approximately correspond to the $BiPyH_2^{2+}$ ion, which is significantly larger than the $KCN$ molecule, one can obtain $G^e_{abs} \sim 10^3 - 10^4$. In some limited situations, when the ion is situated directly on the top of the spike and with the parameters $C_0 \sim 1$, $\beta \sim 1$, $l_1 \sim 100nm$, $a \sim 0.5nm$, $r \sim 0.25nm$ and $B_e \sim 165$, the enhancement coefficient can be $G^e_{abs} \sim 10^9 - 10^{10}$. This result means that for separate molecules, situated in the points with maximum values of the electromagnetic field and its derivatives the enhancement coefficient is huge. Thus the enhancement of the processes of the first type can be very large.

In case we suppose, that the enhancement is determined by the process of the second type only, the enhancement coefficient has the form [7]

$$G^n_{abs} = \frac{(\sigma_{abs})_{surf}}{(\sigma_{abs})_{vol}} \sim C_0^2 \beta^2 \left(\frac{1}{2}B_n\right)^2 \left(\frac{l_1}{r}\right)^{2\beta} \left(\frac{a}{r}\right)^2. \qquad (12)$$

Here we omit estimation of $B_n$, because of its complexity. Some crude estimation is made in [7] for the totally symmetric breathing mode, which demonstrates that $B_n \gg 1$ too. One should note, that for some limited situations, when the molecule is situated directly on the top of the spike, the enhancement coefficient for the processes of the second type is huge also. Thus the enhancement arises for both processes and hence for the cross-section of the infrared absorption.

## Classification of the moments and selection rules

In accordance with our theory the processes of the first type, associated with excitation of the electron shell and further relaxation to the ground state are enhanced due to the quadrupole interaction with the moments $Q_{e,x,x}, Q_{e,y,y}, Q_{e,z,z}$, which have a constant sign. Thus all the quadrupole moments can be divided into two groups:

1. $Q_{e,x,x}, Q_{e,y,y}, Q_{e,z,z}$ main moments with a constant sign and

2. minor moments $Q_{e,x,y}, Q_{e,x,z}, Q_{e,y,z}$ with a changeable sign, which are not responsible for the enhancement.

Moreover, because of the enhancement of the electric field component $E_z$, which is perpendicular to the surface, the moment $d_{e,z}$ also refer to the main moments in case of monolayer coverage of the substrate. Since the tangential components of the field $E_x$ and $E_y$ near the surface are nearly equal to zero, because of the screening by the conduction electrons, the moments $d_x$ and $d_y$ are the minor ones in this case. For multilayer coverage, or solution of the molecules, the molecules orientation with respect to the enhanced electric field can be arbitrary and all the dipole moments refer to the main ones. Further we shall consider the enhancement on molecules with the $D_{2h}$ symmetry group, when all the dipole and quadrupole moments transform after irreducible representations of the symmetry group. The following selection rule is valid for the contributions for the processes of the first type in general case [8],

$$\Gamma_{(s,p)} \in \Gamma_f \quad , \tag{13}$$

where $f$ - is the moment, which is responsible for the relaxation of the electron shell. Thus one can say that the process of absorption of the $(s, p)$ vibration is allowed, in case there is a moment, which transforms after the same irreducible representation as the vibrational mode.

As it is demonstrated in [8] for the processes of the second type, the values $\Delta d_{n,(s,p),i}$ and $\Delta Q_{n,(s,p),i,k}$ follow selection rules, which coincide with (13).

$$\Delta d_{n,(s,p),i} \neq 0 \tag{14}$$

in case

$$\Gamma_{(s,p)} \in \Gamma_{d_{e,i}} \tag{15}$$

and

$$\Delta Q_{n,(s,p),i,k} \neq 0 \tag{16}$$

in case

$$\Gamma_{(s,p)} \in \Gamma_{Q_{e,i,k}} \tag{17}$$

Taking into account estimations of the enhancement for the processes of the first and the second type [7], the following regularities of the enhanced spectra of symmetrical molecules were established in [8].

The most enhancement experiences the line, caused by the breathing totally symmetric vibration. In addition the lines caused by other totally symmetric vibrations can be enhanced also. Since the electric field component $E_z$, which is perpendicular to the surface experiences strong enhancement, then the lines, caused by vibrations, transforming as the $d_{e,z}$ moment experience strong enhancement. In case of multilayer coverage, or a solution, when the orientation of the molecules with respect to the surface is arbitrary, all the dipole moments can be the main ones, since the molecules have all projections of the dipole moment on the enhanced electric field. In this case appearance of the lines, caused by vibrations transforming as any dipole moment is possible.

For the molecules with $D_{2h}$ symmetry group, the above regularities result in appearance of forbidden lines, caused by totally symmetric vibrations, transforming after the unit irreducible representation $A_g$. In addition in case of the multilayer coverage or the solution of the molecules appearance of the lines, caused by vibrations with $B_{1u}, B_{2u}$ and $B_{3u}$ irreducible representations is possible. The lines with $A_u, B_{1g}, B_{2g}, B_{3g}$ irreducible representations are forbidden.

# Analysis of experimental SEIRA spectra of the ion $BiPyH_2^{2+}$ adsorbed on copper

Below we shall analyze the spectra of diprotonated $BiPyH_2^{2+}$ [12], adsorbed on $Cu$, which has the $D_{2h}$ symmetry group. The SEIRA spectra of this ion are presented on figure 1.

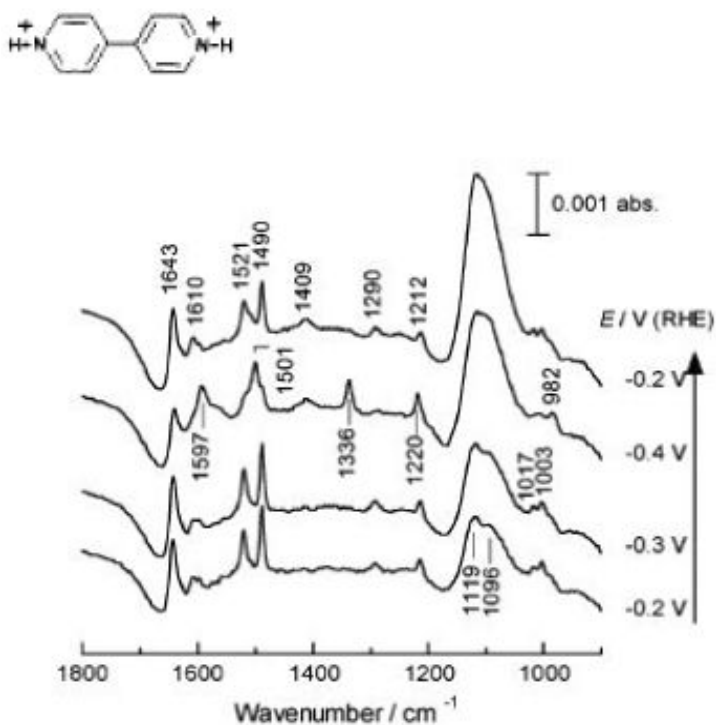

**Figure 1**. SEIRA spectra of $BiPyH_2^{2+}$ for various applied potentials

One should note, that the broad band near $1100 cm^{-1}$ refers to the spectra of water and perchlorate, which are present in the system [12]. In accordance with [12] $BiPyH_2^{2+}$ adsorbs with its plane parallel to the plane of the substrate. The wavenumbers of main lines, which are observed in the spectra of $BiPyH_2^{2+}$ and their assignment are presented in Table 1. One can see strong lines, caused by the totally symmetric vibrations, transforming after the unit irreducible representation $A_g$, which are forbidden in the usual infrared absorption, that corresponds to our theory completely. In addition the authors observe the lines, caused by vibrations with the $B_{1u}$ и $B_{2u}$ irreducible representations. One should note that the assignment to the irreducible representations of the $B_{1u}, B_{2u}, B_{3u}$ type depends on the orientation of the molecule with respect to the coordinate system. Regretfully the authors of [12] do not present this information. However in any case appearance of sufficiently strong lines with the «u» symmetry type is in a good agreement with our theory. Thus analysis of the SEIRA spectra of $BiPyH_2^{2+}$ confirms our point of view.

**Table 1.** Assignment of the SEIRA lines of $BiPyH_2^{2+}$

| The lines of the enhanced absorption of $BiPyH_2^{2+}$. Applied potential is (-0.3-0.1 V) | Assignment of the lines of $BiPyH_2^{2+}$ in the $D_{2h}$ group |
|---|---|
| 1643 (s) | $A_g$ |
| 1610 (w) | $B_{2u}$ |
| 1597 (w) | $B_{2u}$ |
| 1521 (s) | $A_g$ |
| 1490 (s) | $B_{2u}$ |
| 1409 (w) | No assignment |
| 1290 (w) | $B_{2u}$ |
| 1212 (w) | $B_{1u}$ |
| 1017 (w) | $A_g$ |

**Analysis of experimental SEIRA spectra of ethylene adsorbed on copper**

We have described main regularities of the SEIRA spectra, which concern first of all to the molecules, adsorbed on metals with high conductivity, such as silver. In this paragraph we shall analyze the spectra of ethylene, adsorbed on copper. It is necessary to point out that copper is the metal with the absolute value of the imaginary part of the complex dielectric constant, which is larger than the one of silver. This fact causes larger absorption of radiation in the surface area and also results in lesser values of the surface field and its derivatives $\partial E_i / \partial x_i$, for the same mean sizes of the roughness as for silver. This effect should result in a lesser enhancement of the lines, caused by the totally symmetric vibrations, which can be of a comparable value and even less in intensity than the lines, caused by the dipole transitions [13]. The second essential effect is significant shift of the lines, compared with the lines of the free molecules. The last effect is associated with stronger interaction of the molecules with the copper surface and stronger distortion of their geometry and their force field due to adsorption. Since the most enhancement arises in the first layer of adsorbed molecules, the strong chemical interaction with the substrate can result in appearance of two lines instead of one. One line arises from the molecules adsorbed in the first layer and is strongly shifted from the position of the same line of a free molecule, while the second line arises from the molecules adsorbed in the second and upper layers and corresponds approximately to the line of the free molecules. Since the second line arises from the molecules, adsorbed in the upper layers, its intensity can be significantly lower, since the field and its derivatives have significantly lesser values. However in any case the lines, caused by the totally symmetric vibrations can be observed in the spectra. Below we shall consider the spectra of ethylene, adsorbed on copper, obtained in [14] (Figure 2).

As it follows from the results of this work, the lines with wavenumbers 1335 and 1617 $cm^{-1}$, which correspond to the vibrations with the unit irreducible representation $A_g$ are observed in the spectra. The wavenumbers, especially for the first line are very close to the ones of a free molecule. Apparently they arise from the molecules, adsorbed in the second and upper layers since their amplitudes are small and the molecules are situated outside the first layer, where the field and its derivatives have maximum values. Besides there are the lines with the wavenumbers 1276 и 1536 $cm^{-1}$, which are referred by the authors of [14] also to the vibrations with the unit irreducible representation $A_g$. One should note that their wavenumbers are far from the ones of the free molecule that can be associated with their adsorption in the first layer, where the large chemical interaction results in a large frequency shift. In addition their intensities are sufficiently strong that confirms their placement in the first layer. Besides there are the lines with wavenumbers 1436 and 948 $cm^{-1}$, which are referred by the authors of [14] to the vibrations with $B_{1u}$ and $B_{3u}$ irreducible representations. (Here the irreducible representations are written out for the molecule with the $x$ axis, which is perpendicular to the molecule plane and the $z$ axis, which is perpendicular to the C=C bond.) It can be seen, that the lines have large frequency shifts with respect to the ones of a free ethylene molecule, which is associated with large chemical interaction of the molecules with copper. One should note appearance of the lines caused by vibrations with $B_{2g}$ and $B_{3g}$ irreducible representations, which must be forbidden in SEIRA. Their amplitudes are very small. Apparently their appearance is associated with large distortion of the ethylene molecules due to adsorption on copper when all lines became allowed in principle.

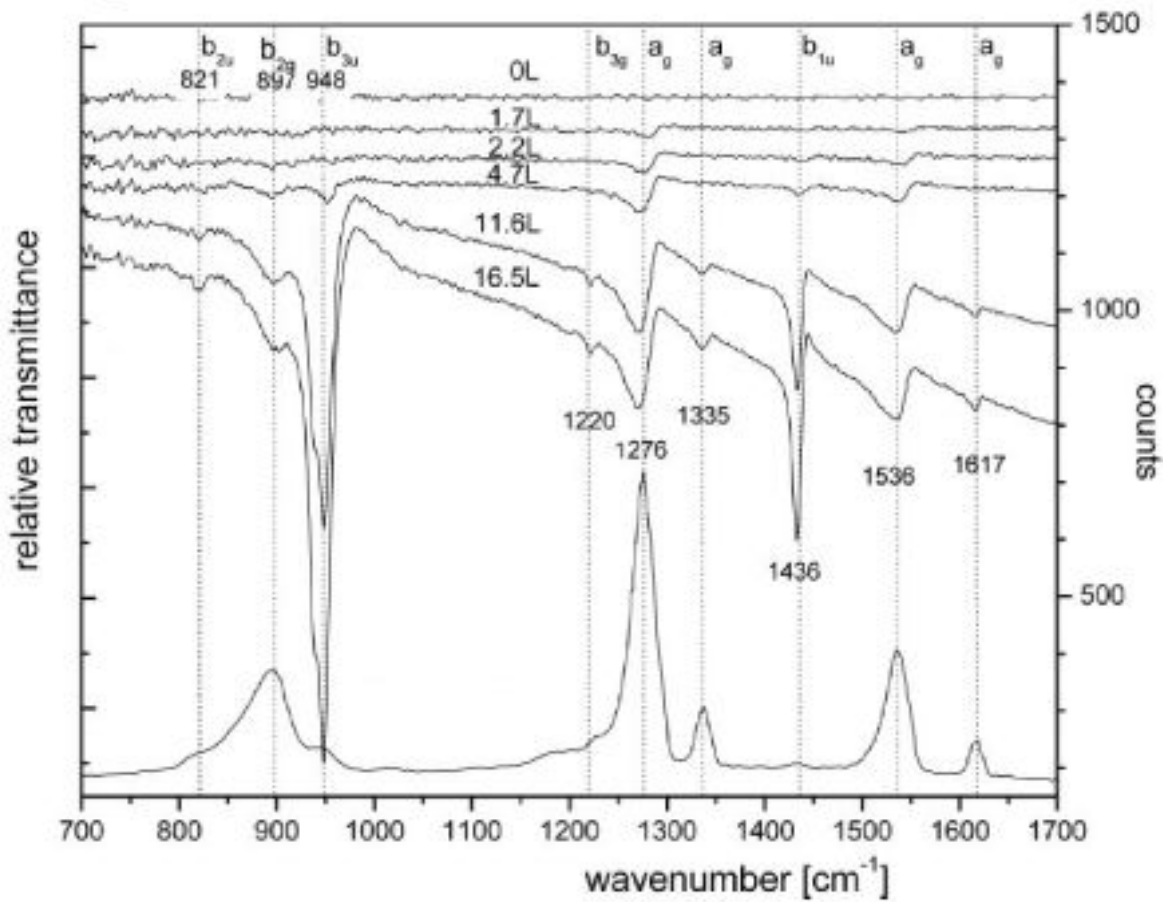

**Figure 2.** SEIRA and SERS spectra of ethylene adsorbed on $Cu$ for various coverage of the substrate

Thus the main peculiarity of the SEIRA spectra of ethylene is appearance of forbidden lines, caused by totally symmetric vibrations transforming after the unit irreducible representation $A_g$. This fact points out the existence of the strong quadrupole light-molecule interaction, which manifests in the SEIRA spectra.

**Appearance of forbidden bands in the infrared spectra of ethylene adsorbed on mordenites**

Another confirmation of the Dipole-Quadrupole theory can be the results published in [15], where the authors observe the lines in the spectral region of the line with the wavenumber 1331 $cm^{-1}$, which corresponds to the line, caused by the totally symmetric vibration with $A_g$ unit irreducible representation. Regretfully we do not know the physical characteristics of the substrates in these experiments. We have in mind the roughness degree and the complex dielectric constants of the mordenites. However appearance of these forbidden lines can be a consequence of manifestation of the strong quadrupole light-molecule interaction in the ethylene molecules, arising due to the electric field with strong space change, arising due to strong irregularity of space near the mordenites surfaces.

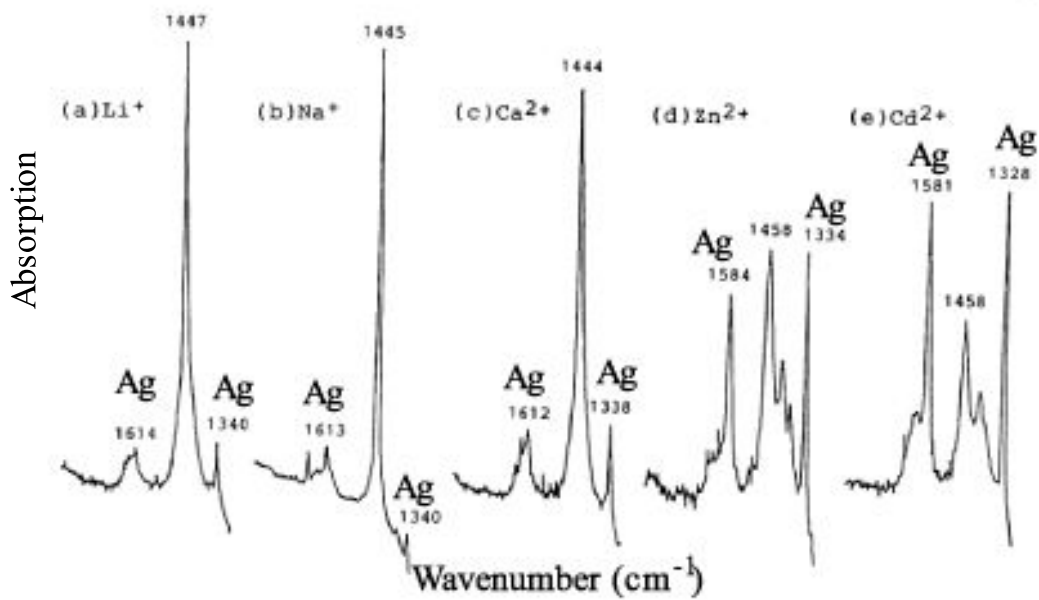

**Figure 3** Surface enhanced infrared absorption spectra of ethylene on mordenites. It is demonstrated appearance of forbidden lines, caused by totally symmetric vibrations with the unit irreducible representation $A_g$.

/